\documentclass[aps,pra,preprint,showpacs]{revtex4-1}

\usepackage{graphicx}
\usepackage{bm}
\usepackage{amsmath}
\usepackage[normalem]{ulem}
\newcommand{\ignore}[1]{}
\newcommand{\bce}{\begin{center}}
\newcommand{\ece}{\end{center}}
\newcommand{\beq}{\begin{equation}}
\newcommand{\eeq}{\end{equation}}
\newcommand{\beqa}{\begin{eqnarray}}
\newcommand{\eeqa}{\end{eqnarray}}
\include{colorrgb}

\begin{document}
\title{Ferromagnetic transition of a two-component Fermi gas of Hard
  Spheres}

\author{F. Arias de Saavedra}
\affiliation {Departamento de F\'\i sica At\'omica, Molecular y Nuclear,  
Universidad de Granada, 18071 Granada, Spain}
\author{F. Mazzanti and J. Boronat}
\affiliation{Departament de F\'{\i}sica i Enginyeria Nuclear,\\
Campus Nord B4-B5, Universitat Polit\`ecnica de Catalunya, 08034
Barcelona, Spain}

\author{A. Polls }
\affiliation{Departament d'Estructura i Constituents de la Mat\`{e}ria
and Institut de Ci\`encies del Cosmos,\\
Universitat de Barcelona, 08028 Barcelona, Spain}


\begin{abstract}
We use microscopic many-body theory to analyze the problem of
itinerant ferromagnetism in a repulsive atomic Fermi gas of Hard
Spheres.  Using simple arguments, we show that the available
theoretical predictions for the onset of the ferromagnetic transition
predict a transition point at a density ($k_F a \sim 1$) that is too
large to be compatible with the universal low-density expansion of the
energy. 
% As a consequence, it is not reasonable to assign the energy of the
% free Fermi sea to the fully polarized phase.
We present new variational calculations for the hard-sphere
Fermi gas, in the framework of Fermi hypperneted chain theory, that
shift the transition to higher densities ($k_F a \sim 1.8$).  Backflow
correlations, which are mainly active in the unpolarized system, are
essential for this shift.
\end{abstract}

\maketitle

\section{Introduction}

Experimental and theoretical progress over the last decade has allowed
for a deeper understanding of phenomena such as superfluidity and
pairing along the BEC-BCS crossover in degenerate Fermi
gases.~\cite{giorgi_rev} In these studies, the diluteness condition of
the system allows for a characterization of the interaction by its
$s$-wave scattering length $a$, a quantity that is easily tuned by
making the system approach a Feshbach resonance. In most cases,
however, the underlying interaction between atoms is short ranged and
attractive, while little attention has been paid to the case of short
range repulsive potentials.

The role played by short-range repulsive interactions has been revived
more recently, with a series of experiments discussing the possible
onset of a phase transition to a ferromagnetic state in two-component
ultracold Fermi gases. In an initial work, Jo \textit{et
  al.}~\cite{jo2009} describe an experiment with $^6$Li atoms which is
supposed to implement the Stoner model~\cite{Stoner33, Huang1987} of
magnetic interactions, and find clear signatures of the ferromagnetic
transition by monitoring the total and kinetic energy of the gas and
its volume. They conclude from these measurements that a transition to
the ferromagnetic state occurs around $x = k_F a \approx 1.9 \pm 0.2$ with
$k_F$ the Fermi momentum of the unpolarized system. In a second and
more recent experiment~\cite{Sanner2011}, the existence of a
ferromagnetic transition in the same system is ruled out, arguing
that very fast molecule formation leads to local heating and to a
pairing instability, destroying the ultracold nature of the gas. They
even conclude that short range repulsive interactions with fermionic
species can possibly not be realized in nature.

The existence of a ferromagnetic phase in dilute, ultracold Fermi
gases interacting through short-range repulsive forces has also been
discussed in recent theoretical works, triggered by the previous
experiments~\cite{jo2009,Sanner2011}.  Positive scattering lengths can
be achieved by moving along the upper branch of a Feshbach resonance
associated to an attractive short-range interaction once a bound state
is formed, or by increasing the range of an overall repulsive
potential. These two mechanisms lead to different physical situations,
as pointed out in Refs.~\cite{Chang2011, Zhou2011}, when a description
based only on the scattering length is not enough.  However, in both
cases it can be argued that when the system is so dilute that the
scattering length and the range of the potential are small compared to
the interparticle distance, only $s$-wave scattering processes are
relevant. Under these conditions, all higher order partial waves can
be discarded in the description of the many-body syste. In any case
and motivated by the previous experiments, the question of whether a
ferromagnetic transition takes place in ultracold Fermi gases has
achieved revived interest recently. However, no consensus has been
reached yet. For instance, in
Ref.~\cite{Sogo2002,Duine2005,LeBlanc2009,Conduit2009} the authors
find that a transition exists in different systems of harmonically
trapped repulsive fermions, while similar conclusions regarding
infinite systems are reported in
Refs.~\cite{Chang2011,Pilati2010,Heiselberg2011}. In other works, the
existence of a transition is either ruled out or argued to depend on
the details of the system analyzed~\cite{Chia2010,Cui2010}.  In this
work, we address this problem again and analyze the possible onset of
a ferromagnetic transition in a system of fermions interacting through
purely repulsive, short ranged forces.  We believe that the question
of weather a system of repulsive fermions can undergo a ferromagnetic
transition is still relevant even though in actual experiments the
high $x$ limit is usually reached by moving along the upper branch of
a Feshbach resonance.

\section{Models}

In the case of a fully polarized Fermi system, $s$-wave
scattering is forbidden by the Pauli principle, and thus the total
energy equals the corresponding kinetic energy of the underlying Fermi
sea if the density is low enough.  In a non-polarized medium, 
$s$-wave scattering between particles
with the same spin orientation is also suppressed but not between
atoms of different spin. All that information can be collected in the
following, effective opposite-spin-channel (OSC) Hamiltonian
\begin{equation} 
H_{\text{OSC}} 
= -\frac{\hbar^2}{2m} \left
( \sum_{i=1}^{N_{\uparrow}} \nabla_i^2 + \sum_{i'=1}^{N_{\downarrow}}
\nabla_{i'}^2 \right ) + \sum_{i,i'}^{N_{\uparrow},N_{\downarrow}}
V(r_{ii'}) \  ,
\label{Hamil1C}
\end{equation}
with $m$ the mass of each of the $N=N_\uparrow+N_\downarrow$ atoms,
$N_\uparrow$ and $N_\downarrow$ being the  number of particles
with spin up and spin down, respectively. 
The Stoner model~\cite{Stoner33, Huang1987}  
corresponds to a mean-field approach to this Hamiltonian 
when the interaction is replaced by a two-body
pseudopotential with a coupling constant proportional to the $s$-wave
scattering length $a$. As commented above, this model 
has been taken as the starting point in recent
theoretical analysis of the ferromagnetic transition.
Contact interactions in standard Monte Carlo simulations of
dilute quantum gases are usually replaced by model potentials of
finite range tuned to reproduce the desired values of scattering
parameters. One of the fully-repulsive potential models 
commonly used to study low-density properties 
of degenerate quantum gases is the hard-sphere interaction,
\begin{equation}
V(r) = \left\{
\begin{array}{cc}
+\infty & r\leq R \\
0       & {\rm otherwise}
\end{array}
\right. \ ,
\label{hardcoreV}
\end{equation}
with an $s$-wave scattering length $a$ equal to the core diameter
$R$, which been used recently in
Refs.~\cite{Chang2011,Pilati2010,Drummond2011}.

When a finite range potential is explored, higher order partial waves
produce contributions that show up when the diluteness condition of
the gas is not fulfilled.  In this case the OSC model, which
explicitly restricts this possibility, shows its limitations.  When
applied to the experiment with $^6$Li atoms of Ref.~\cite{jo2009}, all
models predict a ferromagnetic transition at a point where the system
is clearly not dilute.  With $k_F$ the Fermi momentum of the
paramagnetic phase, the Stoner model predicts the transition at $k_F a
= \pi/2$, while the Monte Carlo simulations of
Refs.~\cite{Chang2011,Pilati2010} lower that prediction to $k_F a
\approx 0.9$, with minor differences between the case of a fully
repulsive interaction and a short-ranged attractive one.  With a
transition point at $k_F a \sim 1$, it is no longer evident that the
OSC model can still be used to describe the system. In this work we
study the possible onset of a ferromagnetic transition in a spin 1/2
Fermi gas of Hard Spheres (i.e.  a two-component Fermi gas) using
existing low-density expansions of the energy per particle and a
variational approach in the framework of the Fermi Hypernetted Chain
Equations (FHNC)~\cite{Rosati1981,polls02}.  As we do not restrict the
analysis to the weakly interacting regime, we allow for the
contribution of scattering in all partial waves by adopting a more
general Hamiltonian of the form
\begin{eqnarray}
H  & = &  -\frac{\hbar^2}{2m} \left ( \sum_{i=1}^{N_{\uparrow}} \nabla_i^2 +
\sum_{i'=1}^{N_{\downarrow}} \nabla_{i'}^2 \right ) 
\nonumber \\
&  &  + 
\sum_{i<j}^{N_{\uparrow}} V(r_{ij}) +
\sum_{i'<j'}^{N_{\downarrow}}V(r_{i'j'}) +
\sum_{i,i'}^{N_{\uparrow},N_{\downarrow}} V(r_{ii'})
\label{Hamiltonian_full}
\end{eqnarray}
where indexes $i,j,..$ and $i',j' ..$ label spin-up and 
spin-down particles, respectively. We stick to the case where the two-body
interaction does not distinguish between different spin
configurations, but explicitly include interactions between spins with
the same orientation, in contrast to the OSC model.
Since we are analyzing the gas of hard-sphere 
fermions, we use~(\ref{hardcoreV}) for $V(r)$ in all three channels
(up-up, down-down and up-down). 

\section{Results}

We are mainly interested in the comparison between the energy per
particle of the paramagnetic, unpolarized system
($N_{\uparrow}=N_{\downarrow}=N/2$) and the ferromagnetic, fully
polarized one ($N_{\uparrow}=N$).
The density determines the Fermi momentum, which is different in each case due to
the spin degeneracy. For the polarized and unpolarized systems one has
$k_F^{\text{P}}=(6\pi^2\rho)^{1/3}$ and 
$k_F^{\text{NP}}=(3\pi^2 \rho)^{1/3}$,
respectively, so $k_F^{\text{P}}=2^{1/3} k_F^{\text{NP}}$. 
As in previous works, we
set $k_F=k_F^{\text{NP}}$ as the momentum scale unit and plot our results in
terms of the dimensionless quantity $x=k_F a$.  Due to the spin
degeneracy, the paramagnetic phase is preferred in the noninteracting
$a \to 0$ limit at zero temperature.  
The stability of the system when the polarization $ \Delta=
(N_{\uparrow} - N_{\downarrow})/N$ increases is determined by the inverse
magnetic susceptibility $\chi^{-1}$,
\begin{equation}
\frac{1}{\chi} = \frac{1}{\rho} \left( \frac{\partial^2 E/N}{\partial
\Delta^2} \right)_{\Delta = 0} \ .
\label{chimag}
\end{equation}
The critical density against spin fluctuations corresponds to the point
where $\chi^{-1}(x_c)=0$. On the other hand, a
sufficient condition  
indicating that the ferromagnetic phase is preferred 
is given by the
criteria $E^{\text{P}}(x_c^\prime) - E^{\text{NP}} (x_c^\prime) = 0$, 
with $E$ the energy of the system. This second estimation, which is the one
we use in the present work, is fulfilled at a
slightly larger density than the first one ($x_c \leq x_c^\prime$). For
instance, in the Stoner model $x_c=\pi / 2 \simeq 1.57$ while $x_c^\prime= (9
\pi /10) \, ( 2^{2/3} -1) \simeq 1.66$.

Several expressions for the ground state energy per particle of a
system of interacting fermions at low densities have been derived in
the past~\cite{efimov65, Bishop1973, Fetter71}. All of them rely on
perturbation theory in terms of a renormalized, effective interaction
obtained from a G-matrix which contains information about the multiple
scattering of particles moving on the correlated medium.
For a strong repulsive potential the massive summation of selected terms is
a convenient strategy   to get a rapidly convergent perturbation series, and
becomes absolutely mandatory for the extreme case of the hard-sphere 
potential due to the divergent character of the matrix elements of the
bare potential.  A very useful expansion 
containing terms beyond $s$-wave contributions is the one derived
by Bishop in Ref.~\cite{Bishop1973}, which is expressed in terms of
the $s$- and $p$-wave scattering lengths $a$ and $a_p$,
respectively, and $r_0$, the $s-$wave effective range, as
\begin{figure}
\begin{center}
\includegraphics*[width=0.45\textwidth]{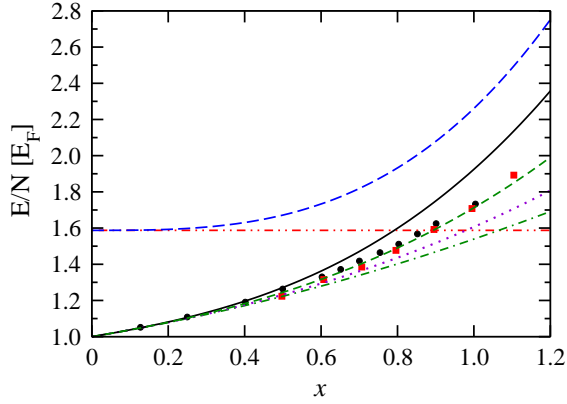}
\caption{(Color online) Energy per particle of the polarized and
  unpolarized hard-sphere Fermi gases. The black solid, violet dotted
  and green dot-dashed lines show the low-density expansion of
  Eq.~(\ref{eqlow}) for the unpolarized system, the same with the
  $p-$wave contribution (proportional to $a_p^3$) removed, and the
  universal prediction of Eq.~(\ref{euniv}). The blue dashed line
  depicts the prediction for the polarized system given in
  Eq.~(\ref{eHS_pol}). The red dot-dot-dashed straight line is the
  constant value assigned to the polarized system in the OSC model,
  while the red squares and solid black circles stand for the DMC
  calculations of Refs.~\cite{Chang2011} and~\cite{Pilati2010} for the
  unpolarized system in the same approximation. }
\label{figStefano}
\end{center}
\end{figure}

\begin{eqnarray}
  \frac{E}{N}  & = & 
  \frac {\hbar^2 k_F^2}{2 m} \left [ \frac {3}{5} + (\nu-1) 
  \left( \frac {2}{3 \pi} (k_F a)  
  + \frac {4(11-2\ln 2)}{35 \pi^2} %(11- 2 \ln 2 )
  (k_F a)^2  \right. \right. 
  \nonumber \\ \label{eqlow}
  & + & \!\!\! 
  \left. \left.  \frac {1}{10 \pi} (k_F r_0) (k_F a)^2 + (0.076 +
  0.057 (\nu -3)) (k_F a)^3 \right)
  \right. 
  \nonumber \\ 
  & + & \!\!\!
  \frac {1}{5 \pi } (\nu +1) (k_F a_p)^3    
  \\   
  & + & \!\!\!
  \left.  
  (\nu-1)(\nu-2) \frac {16}{27 \pi^3} (4 \pi - 3^{3/2})
  (k_F a)^4 \ln (k_F a) + ... \right ]
  \nonumber 
\end{eqnarray}
with $\nu=1(2)$ the spin degeneracy of the polarized (unpolarized)
system.  In this expansion, the $p$-wave scattering length is defined
in terms of the $p$-wave phase shift $\delta_1(k)$ as $a_p=\left.(-k^3
{\rm cot}\delta_1(k)/3)^{-1/3}\right|_{k\to 0}$. For the particular
case of the hard-sphere interaction, where $a_p=a$
and $r_0=2a/3$, the previous expression reduces to the one originally
derived by Efimov~\cite{efimov65}.

For an unpolarized system of Hard Spheres and with $x=k_F a$, the
previous expression reduces to
\begin{equation}
\frac{E^\text{NP}}{N}  = \frac {\hbar^2 k_F^2}{2 m} 
\left[ \frac {3}{5} + \frac {2}{3 \pi}
x + \frac {4(11-2\ln 2)} {35 \pi^2} x^2 + 0.231 x^3 \right]  \ ,
\label{eHS_unpol}
\end{equation}
while for the polarized system one recovers
\begin{equation}
 \frac{E^\text{P}}{N}   = \frac {\hbar^2 k_F^2}{2 m} \left[ 
   \frac{3}{5} + \frac {2}{5 \pi} x^3 \right] \ .  
\label{eHS_pol}
\end{equation}
Notice that in these expressions and up to order $x^3$,
$E^\text{NP}/N$ contains $s$- and $p$-wave contributions besides the
kinetic energy of the corresponding free Fermi sea, while $s$-wave
effects are suppressed in $E^\text{P}/N$. However, there is still a
$p$-wave contribution of order $x^3$ that becomes important when the
density increases. This can be seen in Fig.~\ref{figStefano}, where we
compare the FN-DMC results of Refs.~\cite{Pilati2010} (black circles)
and~\cite{Chang2011} (red squares) for the unpolarized gas to
different instances of Eq.~(\ref{eHS_unpol}) obtained by keeping all
terms (black solid line), removing the $p$-wave contributions (violet
dotted line), or keeping only the {\em universal} terms (green
dot-dashed line), which are of order $x^2$ and depend on the
interaction only through the $s-$wave scattering length
\begin{equation}
 \frac{E^\text{NP}_u}{N} = \frac {\hbar^2 k_F^2}{2 m} \left [\frac {3}{5} + \frac
  {2}{3 \pi} x + \frac {4} {35 \pi^2} (11- 2 \ln 2 ) x^2 \right ] .
\label{euniv}
\end{equation}
To this same order, $E^\text{P}/N$ reduces to the kinetic energy of the
underlaying polarized Fermi sea (red dot-dot-dashed line).  

The previous expansions are asymptotic in nature and so it is
difficult to accurately determine its convergence radius. However, in
the particular case of a hard-sphere interaction, they are known to
reproduce the equation of state up to relatively large values of $x$.
Furthermore, they can not be directly compared to the OSC model as it
is implicitly assumed in their derivation that the Hamiltonian
contains the same interatomic potential in all channels. It is however
possible to obtain a prediction for the OSC model if the leading terms
in the expansion are renormalized accordingly. The OSC model discards
interactions between particles with the same spin orientation, which
in a first approximation can be thought as forming pairs in a triplet
state of total spin $S=1$. The wave function in configuration space
for these pairs is therefore totally antisymmetric and thus $s$-wave
scattering processes between them are suppressed. In this way and to
leading order, the $s$-wave contributions in Eq.~(\ref{eqlow}) should
be the same in the two model Hamiltonians of Eqs.~(\ref{Hamil1C})
and~(\ref{Hamiltonian_full}). For $S=1$ states, $p$-wave scattering is
the leading contribution, but only the symmetric combination of
unparallel spins is allowed in the OSC model. Assuming all three spin
configurations contribute the same to the triplet state in the
Hamiltonian $H$ in Eq.~(\ref{Hamiltonian_full}), we conclude that the
$p$-wave contribution to $H_{OSC}$ should be one third of the $p$-wave
contribution to $H$. Taking all these facts into account, we can build
an expansion of the energy per particle of the OSC model by simply
weighting with an extra factor $1/3$ the term proportional to $a_p^3$
in Eq.~(\ref{eqlow}). Doing so leads to the green dashed line in
Fig.~\ref{figStefano}, which accurately reproduces the Monte Carlo
calculations of Refs.~\cite{Chang2011} and~\cite{Pilati2010} which
were obtained from the model Hamiltonian $H_{OSC}$.  Accordingly, one
may conclude that the low density expansions given above are valid in
the whole range of $x$ values considered, and that both $s$- and
$p$-wave scattering processes contribute significantly to the total
energy per particle when $x$ increases.

Assuming then that the above low density expansions provide a good
description of the energy per particle of the polarized and
unpolarized systems at $x\sim 1$, one can draw several conclusions
from the different curves in Fig.~\ref{figStefano}. For instance, one
sees from the figure that the $x^3$ terms add important contributions
to $E^\text{NP}_u/N$.  Adding only $x^3$ terms to the $s$-wave rises
the energy per particle slightly, but the most significant effect
appears when the $p$-wave contributions are added, bringing the violet
dotted line into the black solid line corresponding to the prediction
for the full model of Eq.~(\ref{Hamiltonian_full}).  Therefore, adding
$p-$wave effects is more important than dressing the $s-$wave
contributions by including the effective range terms.  Furthermore,
the good agreement between the Monte Carlo predictions and the
properly modified low-density expansion for the OSC model, indicates
that while $p$-wave contributions are important, no higher order
partial waves contribute significantly, as these are not included in
the expansion. We thus conclude that $s$- and $p$-wave processes are
the only ones that contribute significantly.

It is also apparent from Fig.~\ref{figStefano} that the universal
behavior, where all predictions yield essentially the same, ceases to
be valid already at $x_0 \sim 0.4$, quite below the value $x\sim 1$
where the simulations based on the 
OSC model predict the transition to
the ferromagnetic phase.  The figure also indicates that the energy
per particle of the polarized gas starts to deviate from the constant
prediction, valid when only $s$-wave scattering is considered, at
about the same value $x=x_0$. Once again, at $x\sim 1$ $p$-wave
contributions introduce significant corrections that can not be
neglected.  We thus conclude that, at least for the hard-sphere Fermi
gas, the low density expansions set the limit of validity of the 
OSC model at $x \sim x_0$.

Knowing the limits of the OSC model, the next step in our discussion is
to analyze the results provided by the full Hamiltonian $H$ of
Eq.~(\ref{Hamiltonian_full}) with the interaction acting in all the
spin channels.  To this end we perform a variational calculation in
the framework of the Fermi Hypernetted Chain equations
(FHNC)~\cite{Rosati1981,polls02,Krotscheck1998}.  Although not exact,
this method has been successfully used in the past to describe
strongly interacting liquids like pure
$^3$He~\cite{Manousakis1983,vivi,buendia}, $^3$He-$^4$He
mixtures~\cite{Fabrocini1982} and nuclear
matter~\cite{fabro88,lova2011}, and is therefore expected to
accurately decribe the physics of the present problem, We use a
variational Slater-Jastrow wave function of the form
\begin{equation}
\Psi_B = \left[ \prod_{i<j} f(r_{ij}) \right] 
D_\uparrow(\vec r_1, \vec r_2, ... \vec r_{N_\uparrow} )
D_\downarrow(\vec r_1', \vec r_2',... \vec r_{N_\downarrow} ) \ ,
\label{eq_jas}
\end{equation}
where $f(r)$ is a two-body correlation factor, while $D_\uparrow$ and
$D_\downarrow$ are Slater determinants of spin-up and spin-down plane
waves filling momentum states up to the Fermi level, and including
backflow correlations~\cite{Feynman1956}.  In our FHNC calculations,
the sum of elementary diagrams is approximated using the interpolating
equation approximation~\cite{vivi,buendia}.  The results obtained
with this wave function~(\ref{eq_jas}) are compared with those
obtained from a less sophisticated version $\Psi$ where backflow
correlations are removed.  In both cases, we take the Jastrow factor
$f(r)$ to be the same for all spin channels, and equal to the solution
of the optimal Euler-Lagrange Hypernetted Chain equations
(HNC/EL)~\cite{Krotscheck1998,Mazzanti2003} for the underlying Bose
gas of atoms of the same mass $m$ interacting through the same
potential and at the same density. We have checked that, by using the
optimal $f(r)$, the energy per particle improves noticeably compared
with other, simpler forms containing few variational parameters,
specially in the case of the unpolarized gas. On the other hand,
backflow correlations enter exponentially in the Slater determinants
of plane waves through the renormalized position
coordinates~\cite{Manousakis1983}
\begin{equation}
{\bf r}_i \to \tilde{\bf r}_i + \lambda \sum_{j\neq i}^N \eta(r_{ij}) 
{\bf r}_{ij}
\label{backflow-1}
\end{equation}
where
\begin{equation}
\eta(r) = \exp\left[ -\left( {\frac{r - r_0}{\omega_0} } \right)^2
  \right]
\label{backflow-2}
\end{equation}
with $\lambda, r_0$ and $\omega_0$ variational parameters that are
optimized at each density. Backflow correlations are capital in an
accurate description of the energetics of strongly correlated systems
like $^3$He,~\cite{carlson,boro} and have been reported to have also a
non-negligible impact in the problem considered here at the densities
where the ferromagnetic transition is predicted~\cite{Chang2011}.

\begin{figure}
\begin{center}
\includegraphics*[width=0.5\textwidth]{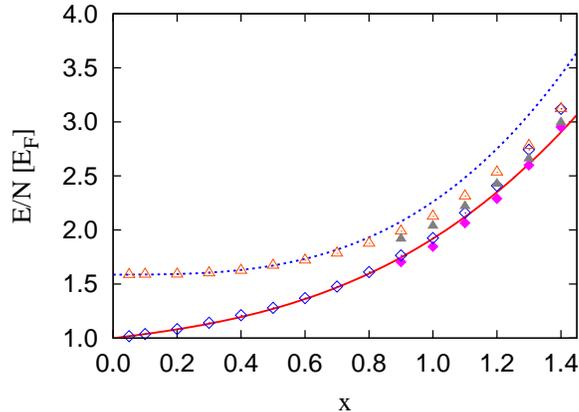}
\caption{(Color online) Energy per particle of the paramagnetic and
  ferromagnetic phases in different approximations. Solid and open
  diamonds: FHNC results for the unpolarized gas including or not
  backflow correlations, respectively. Solid and open triangles: same
  for the polarized system. The solid and dotted lines are the
  predictions of the expansions in Eqs.~(\ref{eHS_unpol})
  and~(\ref{eHS_pol}), respectively.  }
\label{figFHNC}
\end{center}
\end{figure}

Figure~\ref{figFHNC} shows our results for the total energy per
particle of the polarized and unpolarized systems in different
approximations. The solid and dotted lines show the low density
expansions of Eqs.~(\ref{eHS_unpol}) and~(\ref{eHS_pol}),
respectively.  The open triangles and diamonds correspond to the
predictions for the polarized and unpolarized systems using the
HNC/EL-optimized wave function of Eq.~(\ref{eq_jas}) without backflow
correlations, while the solid symbols stand for the same predictions
including backflow.  Several conclusions can be drawn from this
figure. On the one hand, it is remarkable the fact that while the low
density expansion for the unpolarized system is everywhere close to
the full many-body calculations, this is not the case for the
polarized system, which overestimates the total energy per particle of
the gas when $x\geq 0.6$.  On the other, while the low density
expansion predictions do not cross in the range of densities analyzed,
our many-body calculations show that the energy per particle of the
polarized system decreases at a higher rate than the corresponding one
for the unpolarized gas.  It is also remarkable to notice that the
total energy per particle of both systems noticeably depends on the
quality of the wave function employed.  As expected, the one including
backflow correlations provides lower variational estimations.  In any
case, the inclusion of backflow correlations not only lowers the
energies, but also pushes the ferromagnetic transition to a higher
density, identified by the point in the figure where the energy of the
paramagnetic state equals that of the ferromagnetic phase (a
sufficient condition that determines a point $x_c^\prime$ where the
ferromagnetic phase is preferred). The open symbols show that $\Psi$
predicts a transition at $x_c^\prime\sim 1.4$, while the solid
symbols, corresponding to $\Psi_B$, approach each other but do not
cross yet in the range of densities considered in the figure. In any
case, it is relevant to realize that the use of a Hamiltonian allowing
the interaction in all three spin channels (up-up, down-down and
up-down), together with a rich variational wave function, bring the
transition point back from $x_c^\prime \sim 1$ to higher values closer
to the prediction of the Stoner model, $x_c^\prime \sim 1.66$.

\begin{figure}
\begin{center}
\includegraphics*[width=0.5\textwidth]{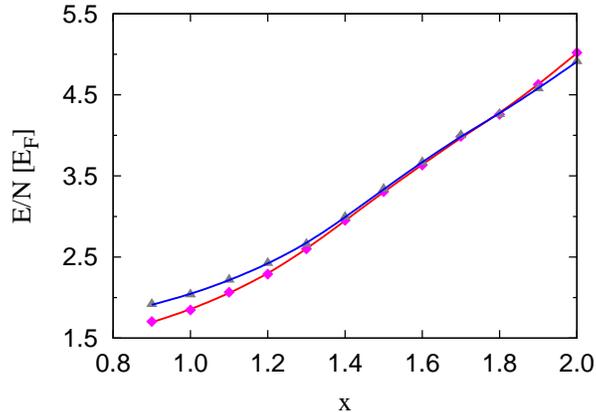}
\caption{(Color online) Energy per particle of the polarized (grey
  triangles) and unpolarized (red diamonds) system obtained from the
  HNC/EL-optimized wave function of Eq.~(\ref{eq_jas}) including
  backsflow correlations, in the framework of the FHNC theory. The
  solid and dotted lines are a guide to the eye.
}
\label{figBackflow}
\end{center}
\end{figure}

The precise density at which our fully correlated model predicts the
transition to the ferromagnetic state can be determined by extending
the range of $x$ values considered.  In Fig.~\ref{figBackflow}
we show our predictions for $E^\text{P}/N$ and $E^\text{NP}/N$ obtained from 
the optimized
wave function $\Psi_B$ of Eq.~(\ref{eq_jas}) including backflow
correlations. As it can be seen, there is a wide range of densities
where $E^\text{P}/N$ and $E^\text{NP}/N$ are fairly close to each other, 
and thus establishing a clear prediction of the transition point is delicate.
Still, we clearly see that for $x > 1.8$ the polarized gas has lower
energy than the unpolarized one, showing that at this point (and
higher densities) the ferromagnetic phase is preferred.  To the present
accuracy we find that both curves cross at $x_c^\prime \sim 1.8$, which
establishes our prediction of the point where the ferromagnetic
transition takes place for the hard-sphere Fermi gas.
This transition occurs at a density that is close to the estimated
freezing density of quantum hard spheres ($x_{\text f} \simeq
1.95$)~\cite{kalos} but slightly below.
Our result for $x_c^\prime$ turns out to be surprisingly close
to the original predictions given in Ref.~\cite{jo2009}, although
recent experiments~\cite{Sanner2011} with $^6$Li atoms seem to
indicate that in nature a pairing instability breaks the
ultracold nature of the gas before the transition takes place.
In any case and as stated above, one should recall that the
mechanisms explored here are related but not equal to those found in
the experiments which exploit the physical properties of the system
close to a resonance. 

\section{Summary and Conclussions}

To summarize, in this work we have analyzed the possible onset of a
phase transition to a ferromagnetic phase for a two-component Fermi
gas interacting through a repulsive hard sphere potential. Motivated
by recent experimental and theoretical works, we have compared our
results with existing predictions~\cite{Chang2011,Pilati2010} for a
simplified Hamiltonian where only correlations between different spins
are allowed. We have shown that, when properly renormalized to
correctly account for $p$-wave scattering processes, already existing
low-density expansions accurately reproduce Monte Carlo simulations of
this simplified model. Based on that, we have shown that $p$-wave
terms contribute significantly to the total energy per particle of the
polarized and unpolarized systems at $k_F a\sim 1$, and that therefore
the simplified Hamiltonian for a fully repulsive interaction
can not be used in this range. Finally, we
have also presented calculations in the framework of the FHNC
equations for a wave function including backflow correlations and an
optimized Jastrow factor, showing that a ferromagnetic transition
takes place at $k_F a \sim 1.8$, surprisingly close to the original
prediction of Ref.~\cite{jo2009}.

\begin{acknowledgments}

This work has been partially supported by Grants 
No.~FIS2009-07390 and~FIS2008-04403, 
and FIS2008-01661 from MICINN (Spain) and FEDER and Grants No.~2009-SGR1003 and 
2009SGR-1289 from the Generalitat de Catalunya (Spain).

\end{acknowledgments}

%\clearpage

\end{document}